\pdfoutput=1

\documentclass[sigconf,screen]{acmart}
\AtBeginDocument{%
  \providecommand\BibTeX{{%
    \normalfont B\kern-0.5em{\scshape i\kern-0.25em b}\kern-0.8em\TeX}}}

\setcopyright{acmlicensed}
\copyrightyear{2024}
\acmYear{2024}
\acmDOI{XXXXXXX.XXXXXXX}

\acmConference[FSE '24]{International Conference on the Foundations of Software Engineering}{Mon 15 - Fri 19 July 2024}{Porto de Galinhas, Brazil, Brazil}
\acmBooktitle{Proceedings of the 32nd ACM Symposium on the Foundations of Software Engineering (FSE '24), November 15--19, 2024, Porto de Galinhas, Brazil}
\acmISBN{978-1-4503-XXXX-X/18/06}

\usepackage{tabularx}

\begin{document}

\title{Automated Unit Test Improvement using Large Language Models at Meta}


\author{Nadia Alshahwan}
\authornote{Author order is alphabetical. The corresponding author is Mark Harman. }
\orcid{https://orcid.org/0009-0009-4763-0396}
\author{Jubin Chheda}
\orcid{https://orcid.org/0009-0005-0311-7890}
\author{Anastasia Finegenova}
\orcid{https://orcid.org/0009-0005-5824-5179}
\author{Beliz Gokkaya}
\orcid{https://orcid.org/0009-0003-0197-6806}
\author{Mark Harman}
\orcid{https://orcid.org/0000-0002-5864-4488}
\author{Inna Harper}
\orcid{https://orcid.org/0009-0008-9359-0949}
\author{Alexandru Marginean}
\orcid{https://orcid.org/0009-0001-5311-762X}
\author{Shubho Sengupta}
\orcid{https://orcid.org/0009-0007-4204-5185}
\author{Eddy Wang}
\orcid{https://orcid.org/0009-0009-8825-6986}
\affiliation{%
  \institution{Meta Platforms Inc.,}
  \streetaddress{1 Hacker Way}
  \city{Menlo Park}
  \state{California}
  \country{USA}
}

\renewcommand{\shortauthors}{Alshawan and Harman, et al.}

\begin{abstract}
This paper describes Meta's TestGen-LLM tool, which uses LLMs to automatically improve existing human-written tests. 
TestGen-LLM verifies that its generated test classes successfully clear a set of filters that assure measurable improvement over the original test suite, thereby eliminating problems due to LLM hallucination.
We describe the deployment of TestGen-LLM  at Meta test-a-thons for the Instagram and Facebook platforms. 
In an evaluation on Reels and Stories products for Instagram, 
75\% of TestGen-LLM's  test cases built correctly, 57\% passed reliably, and 25\% increased coverage.
During Meta's Instagram and Facebook test-a-thons, it improved 11.5\% of all classes to which it was applied, with 73\%  of its recommendations being accepted for production deployment by Meta software engineers.
We believe this is the first report on industrial scale deployment of LLM-generated code backed by such assurances of code improvement.
\end{abstract}




\maketitle

\section{Introduction}
As part of our overall mission to automate unit test generation for Android code, we have developed an automated test class improver, TestGen-LLM.
TestGen-LLM uses two of Meta's\footnote{The two LLMs used by TestGen-LLM were constructed at Meta for general purpose internal use, but they are not the focus of this paper, which is about an LLM-agnostic ensemble approach, its application to test class improvement at Meta and our experience with it at Meta's Test-a-thons. 
Because details are commercially sensitive (and not relevant to this paper), we do not give details of the two LLMs, simply calling them `LLM1' and `LLM2' in this paper. } Large Language Models (LLMs) to extend existing, human-written, Kotlin test classes by generating additional test cases that cover previously missed corner cases, and that increase overall test coverage.
TestGen-LLM is an example of Assured Offline LLM-Based Software Engineering (Assured Offline LLMSE) \cite{mhetal:intense24-keynote}.

That is, unlike other LLM-based code and test generation techniques, TestGen–LLM uses Assured Offline LLMSE to embed the language models, as a service, in a larger software engineering workflow that ultimately recommends fully formed software  improvements rather than smaller code snippets.
These fully-formed code improvements are backed by verifiable guarantees for improvement and non-regression of existing behavior.
A filtration process discards any test case that cannot be guaranteed to meet the assurances.

The filtration process can be used to evaluate the performance of a particular LLM, prompt strategy, or choice of hyper-parameters. 
For this reason, we include telemetry to log the behavior of every execution so that we can evaluate different choices.
However, the same infrastructure can also be used as a kind of ensemble learning approach to find test class improvement recommendations.
TestGen-LLM thus has two use cases:

\begin{enumerate}
\item {\bf Evaluation}: To evaluate the effects of different LLMs, prompting strategies,  and hyper-parameters  on the automatically measurable and verifiable improvements they make to existing code.
\item {\bf Deployment}: To fully automate human-independent test class improvement, using a collection of LLMs, prompting strategies, and hyper-parameters to automatically produce code improvement recommendations that are backed by  
\begin{enumerate}
    \item Detailed automatically-generated documentation that measures the improvement achieved by the new version of the test class;
    \item Verifiable guarantees that the recommended test class does not regress any important properties of the existing version of the test class.
\end{enumerate}
\end{enumerate}

TestGen-LLM has been used in both of these modes.
The evaluation mode was used as a prelude to deployment, allowing us to investigate and tune such choices of LLM, prompt strategy and temperature.
It was also used after initial deployment to tune parameters for the subsequent, more widespread, release of the tool to engineers at Meta.
The evaluation mode also allows us to report findings for evaluation (See Section~\ref{sec:eval}).

Having arrived at sensible parameter choices, based on evaluation, TestGen-LLM was  used at Meta to support engineers in various test improving activities, such as test-a-thons, in which a focused team of engineers target a particular aspect of one of Meta's products, in order to enhance existing testing.

Initial planning for TestGen-LLM took place in spring 2023 \cite{alshahwan:software}, with initial development in summer and autumn, and evaluation and onward optimization through winter 2023.
This paper describes TestGen-LLM and reports our experience in developing and deploying it, through these test-a-thons for Instagram and Facebook.
The primary contributions of the paper are:

\begin{enumerate}
    \item The introduction of the first example of Assured LLM-based Software Engineering (Assured LLMSE) \cite{mhetal:intense24-keynote}. 
    In particular, we believe this is the first paper to report on LLM-generated code that has been developed independent of human intervention (other than final review sign off), and landed into large scale industrial production systems with guaranteed assurances for improvement over the existing code base.
    \item In an evaluation on Reels and Stories products for Instagram, 75\% of TestGen-LLM  test cases generated  built correctly, 57\% passed reliably, and 25\% increased coverage.
    \item A report on the qualitative and quantitative results of development, deployment and evolution at Meta in 2023. When deployed to incrementally improve the coverage of production test classes on Instagram and Facebook more generally, 
    TestGen-LLM was able to improve 10\% of all classes to which it was applied and 73\% of its test improvements were accepted by developers, and landed into production.
    \item A description of the lessons learned,  open problems and research challenges raised by this application of Assured LLMSE to software test improvement. 
    
\end{enumerate}

\section{The TestGen-LLM System}
\begin{figure*}[t]
\centerline{\includegraphics[width=\linewidth]{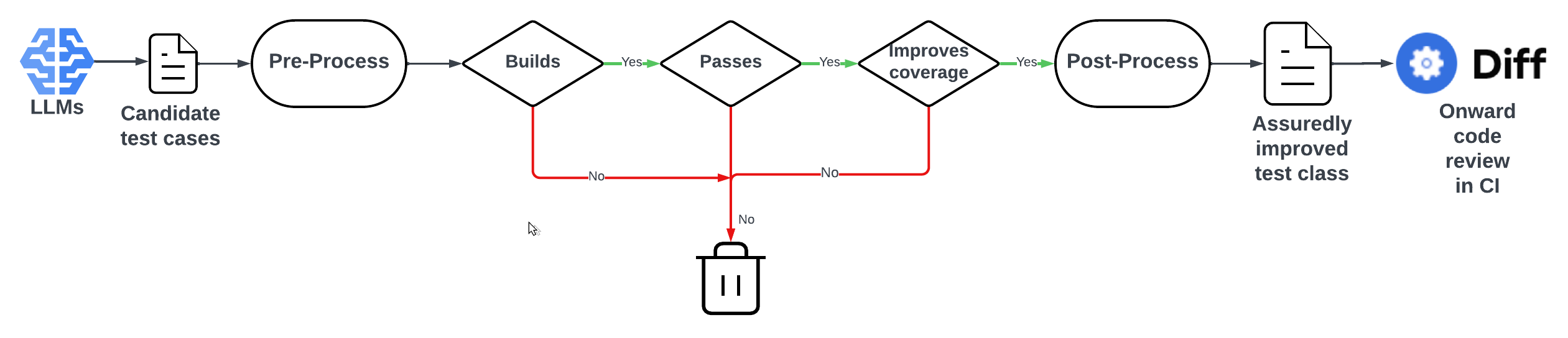}}
\vspace{-3mm}
\caption{TestGen-LLM top level architecture (an instance of Assured Offline LLMSE \cite{mhetal:intense24-keynote}).
\label{fig:top_level}}
\end{figure*}

TestGen-LLM achieves the assurances it offers to code reviewers by applying a series of progressively demanding semantic filters to candidate solutions generated by the language models.
Figure~\ref{fig:top_level} depicts the top level architecture of the TestGen-LLM system.

The first filter simply checks that the candidate code is fully buildable within the existing infrastructure of the app under test. 
Any code that does not build is immediately discarded and thereby removed from further consideration.

The second filter executes the generated tests, which build by definition, due to their clearing the first filter.
Any test that does not pass may reveal a fault. 
However, it is  more likely that the test simply contains an incorrect assertion. 
We want an entirely automated workflow.
Without an automatable test oracle \cite{ebetal:oracle}, TestGen-LLM cannot automatically determine whether a failing test has found a bug, or whether it merely contains an incorrect test assertion.
Therefore, TestGen-LLM discards any test case that does not pass on first execution. 
The effect of this filter is to preserve only tests that can be used for regression testing \cite{symh:regression-survey}. 
Such tests, since they pass, protect the existing functionality of the system under test against future regressions.

A test that passes on first execution, may only coincidentally pass on the first occasion on which it is executed. 
More generally, a  test that passes on some occasions and fails on others, when executed in an entirely identical environment, is called a `flaky' test \cite{luo:flaky}.
Flaky tests are one of the most significant problems for industrial software testing \cite{mhpoh:scam18-keynote}, and so we clearly do not want TestGen-LLM to introduce flakiness.
TestGen-LLM thus uses the `passes' filter to discard any flaky tests.
The filter uses the simple, widely-used and effective approach of repeated execution \cite{mcetal:flakime,luo:flaky}; a test that does not pass on every one of five executions is deemed flaky.
Since the generated test cases  are unit level tests, repeated execution is a relatively computationally inexpensive approach to filtering out flaky tests.

A candidate test case that passes through the first two filters is guaranteed to provide reliable signal for regression testing. 
However, it may simply repeat  the test behaviour of one of the existing tests. 
Such duplication of test effort would be a waste of resources. 
Furthermore, there will be no meaningful way in which TestGen-LLM could reliably {\em claim} that the original test class had been {\em improved} in some {\em measurable} way.
Therefore, the final filter applied to non-flaky passing tests measures their coverage. 
Any test that does not improve coverage is also discarded.

The candidate test cases that pass through all three filters are thus guaranteed to improve the existing test class, and to provide reliable regression test signal.
The pre- and post-processing are steps used to extract test cases and re-construct test classes.

\subsection{Advantages of code improvement style LLM code generation}

TestGen-LLM uses an approach called `Assured LLM-based Software Engineering' (Assured LLMSE) \cite{mhetal:intense24-keynote}.
In the remainder of this section, we set the principal advantages we have experienced from our development and deployment of TestGen-LLM,
focusing on those we believe likely to  carry over to other Assured LLMSE applications, and are not confined merely to test generation.

\subsubsection{Measurable improvement} 
At Meta, a code change submitted to the Continuous Integration (CI) system  is called a `diff' (short for differential).
The diffs that TestGen-LLM generates include a precise measurement of the improvement they offer. 
For testing, the goal is to increase coverage, especially of corner cases that might have been missed. 
The diffs submitted by TestGen-LLM document their own coverage improvement to support the claim that they improve the code base.

\subsubsection{Verifiable guarantees of non-regression} 
The diffs also include evidence of how they guard against regression. 
For test generation, TestGen-LLM simply augments an existing test class with additional test cases, retaining all existing test cases and thereby guaranteeing there will be no regression, by construction. 

Assured LLMSE can also target operational characteristics, such as performance, for which it will necessarily rely on tests to guard against regression \cite{mhetal:intense24-keynote}. 
This was the motivation for our initial  focus on test generation: once we have reliable automated regression testing at scale, we can use Assured LLMSE to target operational code properties, such as performance.

\subsubsection{Ensemble approach}
Different LLMs have different strengths. 
Even the same LLM can produce multiple candidate solutions for a given prompt. 
As our results show, although some prompts, parameters and underlying LLM technologies perform better than others for a given test class, 
each combination tends to contribute {\em uniquely} to the overall number of test cases found (See Section~\ref{sec:experiment}).
It is therefore highly advantageous to formulate the problem in such a way that it is amenable to an ensemble-style learning approach \cite{dong:ensemble-survey}. 
In such an ensemble approach, the best aspects of many LLMs, their prompts and parameters, can also be combined to give an overall improvement recommendation. 

The LLM produces code components, not entire programs. 
Components are composable, and therefore can be provided by multiple different LLMs, working as an ensemble. 
For the test improvement instance of Assured LLMSE, each component is a test case.
Test cases compose very naturally to form a test classes and test suites.  

In the more general case of Assured LLMSE \cite{mhetal:intense24-keynote}, LLM recommendations can be defined as code modifications
as they typically are for automated repair \cite{legoues:cacm-survey,legoues:genprog-sqj,huang:repair-survey,Repairnator,ametal:sapfix}. 
Code modifications are also composable, and thereby support the overall ensemble approach to Assured LLMSE.

\subsubsection{Helps humans; does not replace them} 
By seeking to improve existing code and tests, rather than constructing them from scratch, TestGen-LLM  works in concert with human engineering effort, insights and domain expertise; it does not seek to replace them.
TestGen-LLM builds on the positive aspects of language models, while simultaneously sidestepping two of the most pressing concerns, raised in the literature and wider community: 

\begin{itemize}
    \item Whether language models will replace human coders (TestGen-LLM is a support {\em not a replacement} for humans).
    \item Whether language model results can be relied upon (TestGen-LLM results come with guarantees that ensure they {\em can} be relied upon).
\end{itemize}

More specifically, there has been much discussion on whether language models will replace human coders, with many arguments on all sides, and no shortage of opinions \cite{mhetal:LLM-survey}.
A debate also continues concerning whether LLMs are beneficial or harmful to education, for example, with arguments against ~\cite{meckler:alert}, and broadly in favor ~\cite{heaven:chat} of their use as educational tools, or as ways to support existing human-led education efforts \cite{sarsa_automatic_2022,macneil_experiences_2023,noever_chatbots_nodate}.
However, since TestGen-LLM is designed to be used as a support tool (to help engineers rather than to replace them), it is unnecessary for us to enter into this discussion in this paper.
Rather, TestGen-LLM's goal is to provide a {\em recommender} system \cite{adomavicius:recommender}, that leaves the software engineer in ultimate control of the code that lands into the code base. 
This also ensures proper engineering oversight and accountability.

There has also been much discussion of the problems of relying on machine-generated code, especially the problem of hallucination \cite{ma:scope,mhetal:LLM-survey}.
The TestGen-LLM approach in particular (and the overall Assured LLMSE approach more generally),  
overcomes hallucination by providing automated verifiable guarantees about the semantic properties of the code it recommends. 
These guarantees mean that the language model itself plays its own role in self accountability, providing at least as strong a semantic guarantee as many human engineers for the recommendations it makes.

\subsubsection{Caters well to LLMs with limited token window size}
The specific formulation we used for test generation (as the problem of extending an existing test class), means that the test generation approach can be effective, even with a very small token window size. 
This is because the test class is typically much smaller than the class under test. 
TestGen-LLM can and does use prompts that provide both the test class and the class under test. 

Providing the class under test does produce better results (as one would expect, since Retrieval Augmented Generation is known to perform well \cite{mhetal:LLM-survey}). 
Nevertheless, we have also been able to successfully extend test classes by providing solely the existing test class, and omitting the class under test and any other details from the prompt.
As our results show, prompts that use solely the test class (and not the class under test) as part of the prompt context, 
can still find additional tests.
Moreover, such prompts found {\em unique} tests not found by other prompts
(see results Section~\ref{sec:extend-test} for more details).

\section{TestGen-LLM Deployment}
In a large company like Meta, it is typically not possible to simply switch on a new technology once developed, and apply it at scale.
First, we must perform initial trials, and then cautiously deploy a Minimal Viable Product (MVP), 
in a well-controlled environment that allows us to gain experience,
before migrating the technology to full deployment.
If we simply deploy an MVP without first gaining this experience, 
then  relatively small issues in the behavior of the technology can become considerably magnified at scale. 
The magnifying effect of small inefficiencies, errors or overlooked details, force multiply at this kind of scale.
To give a sense of the scale, note that the central repository receives well over 100,000 commits per week \cite{mhpoh:scam18-keynote},
while the apps under test have client-side code bases of tens of millions of lines of code each, communicating with back-end server-side infrastructure of hundreds of millions of lines of code.

The deployment workflow thus follows a gradual incremental deployment plan, in which the MVP is gradually evolved and matured from proof of concept to deployed tool/infrastructure, over a series of increasingly larger-scale, and  increasingly less tightly constrained trials.
In the initial phases, the proto-MVP is applied at a very small scale, and in a tightly-controlled environment. 
In the later stages, after working in changes from the feedback on these initial deployments, the MVP becomes a more fully-fledged software engineering tool and is deployed freestanding (without detailed human oversight).
In this section we describe the process by which we migrated from initial proof of concept to deployment.

\subsection{Initial Trial}

As an initial trial, we used an initial version of TestGen-LLM to create eight diffs and submitted these into Meta's standard continuous integration code review system.

The initial trial re-enforced the 
importance of giving individual guarantees per test case, and the value of  the improvement assurances.
The engineers who reviewed the initial diffs reported that the following  two additional features would maximize the ease and speed with which they could review the recommendations from TestGen-LLM:

\noindent
{\bf
1. The importance of individual test level guarantees.}
In the initial MVP, 
we had only implemented class-level improvement guarantees. 
That is, the overall test class is improved, but there was no guarantee that each {\em individual} test case contributes to this improvement.
For example, one of the diffs that ultimately landed initially contained four test cases. 
However, all four were only superficially different. 
By measuring the individual coverage contributed by each test case, the updated version of TestGen-LLM now automatically weeds out such duplicated test effort. 
This `per test case' approach (rather than `per test class')  is slower to compute, but it gives TestGen-LLM the overall ability to more easily mix and match the results from different LLMs and different responses from the same LLM; the ensemble-style approach.

\noindent
{\bf
2. Useful to give more coverage details:}
The TestGen-LLM initial MVP reported only the files for which the improvement suggestion achieves extra coverage. 
Some  of the engineers who reviewed the diffs asked whether the tool could provide full specific coverage information for each file, compared to the original. 
The TestGen-LLM MVP was therefore updated to report all details of the coverage achieved.

In this way we are able to gain valuable insights on the initial version of the tooling before deploying at wider scale. 
Based on the lessons learnt from this initial trail, we went on to develop a new version of TestGen-LLM, which was used as part of the next available test-a-thon exercise, in which engineers specifically sought to write new tests and extend existing test cases. 
We describe this next phase of deployment in the next section.

\subsection{Instagram Test-a-thon November 2023}

We used the the second version of the TestGen-LLM MVP  to generate extra tests as part of the Instagram test-a-thon, 
which ran from November 15th - 20th 2023. 
Test-a-thons are regular human-centric activities in which engineers seek allocate specific focused time to writing test cases.
In the  November Instagram test-a-thon, TestGen-LLM was used in a carefully managed way in order to assess its suitability for wider and less closely-managed use.

\noindent
{\bf 
Initial calibration:} 
The engineer leading the test-a-thon first identified an initial component of interest for her team, 
so that she could  confirm (or refute) that the diffs generated by the TestGen-LLM  MVP would be suitable for consideration. 
TestGen-LLM  produced three diffs for this component, each with a single test class  extension for one of three different sub-components. 
In all three cases, the test class improvements were deemed acceptable.
It was  found that the verifiable claims for improvement made it easy to accept and land these tests into production.
The engineer also greatly appreciated the way in which the generated tests' {\em coding style} closely mimicked that of the existing human–written test classes for each sub-component.
Therefore, based on this initial trial,
it was agreed to deploy TestGen-LLM  as part of the test-a-thon.

\noindent
{\bf 
Identifying the test classes to be targeted:}
During the three days of the test-a-thon, 36 engineers spent significant portions of their time focused on writing additional test cases for specific Instagram products targeted by the test-a-thon.
The test classes and products chosen for the test-a-thon were those that had been the subject of recent intensive re-factoring activity.

Whenever a diff was submitted by an engineer as part of  the test-a-thon process between 15th and 18th November 2023, 
TestGen-LLM was executed on the directory in which the test class resided, 
seeking to extend any of the test classes that reside in that directory.
There is a generally close correspondence between directories and sub-components.
In particular, all test classes in the directory are typically  built using the same  build rule and, therefore, TestGen-LLM thus automatically measures the additional coverage achieved over and above all the existing test classes residing in the same directory as the human-written test class.

\noindent
{\bf
Simulating a diff time deployment:} 
There are two primary modes in which automated testing can interpose in continuous integration, which we typically call `diff' time and `post-land' time \cite{mhetal:ssbse18-keynote,jaetal:mia}.
At diff time, the tests are recommended to the engineer at the time they submit a related diff, whereas at post-land time, the test is recommended to the engineer at some arbitrary point after they have landed the relevant diff.

In previous work on deployment of automated testing technology at Meta, we have repeatedly found that diff time deployment is far more effective, because it maximizes relevance \cite{mhpoh:scam18-keynote}.
When a test is recommended at diff time, the engineer concerned already has the full context of the existing testing in place, and the code under test.
As such, the engineer is in a much better position to quickly and correctly assess the recommended test. 
This increased relevance typically maximizes the impact of the test recommender system and the signals it provides. 
For this reason, we seek to prioritize diff time deployment, wherever possible.

Through the test-a-thon, we were able to gain experience of diff-time deployment mode, and thereby gain insights and experience on how this technology would play out when deployed in this mode.
Although the potential reviewers had technically already landed {\em some} of the diffs containing test classes extended by TestGen-LLM, 
they were much more likely to have context, and to have only very {\em recently} landed test classes in the same directory. 
However, we cannot claim that this was a {\em perfect} example of diff time deployment: TestGen-LLM might also have benefited from the fact that this was a test-a-thon, and therefore, there was a greater-than-usual focus on testing and the context in which the tests were being deployed.

\noindent
{\bf
How the TestGen-LLM diffs were constructed for the November Instagram Test-a-thon:} 
We constructed the diff summaries and test plans and submitted them manually, but used only information computed directly by TestGen-LLM.  Where we included any text in the summary indicating our own interpretation of the results, this was contained in a blue-box “Note” to distinguish this text as human-generated and not machine-generated.  
In the second Instagram test-a-thon (see Section~\ref{sec:igtestathon2}) we fully automated the construction of the content and claims made by TestGen-LLM in diff comments and test plans.

\subsubsection{Outcomes from the first Instagram Test-a-thon}
\label{sec:cheeky-todo}
During the first test-a-thon, 36 engineers landed 105 unit test diffs, of which 16 were generated by TestGen-LLM.
In total, TestGen-LLM created 17 diffs.
One diff was abandoned because the test case did not include any assertion.
16 landed into production.
The test case in the rejected diff attempted to test a partially implemented function, for which the LLM code left a comment indicating that an assertion was to be added as a “TODO”.
The test case did extend coverage (simply by executing the previously unexecuted method-under-test), but it was rejected by the engineer reviewing the diff because it failed to contain an assertion.

The largest coverage improvement was achieved by a TestGen-LLM diff that covered a method not previously covered, and thus generated a lot of additional coverage, including:
\begin{itemize}
\item 28 new files covered that were not previously covered.
\item 13 files which were previously partially covered, but for which the improved test class extended coverage
\item 3 A/B testing gate keepers (for which generated A/B decision making code was additionally covered).
\end{itemize}

The smallest coverage improvement was produced by a diff that covered a single additional line (an early {\tt return}) in a file that was already partially covered by the existing test classes.

Anonymized results for top 10 performers among the 36 engineers at the the test-a-thon are shown in Table~\ref{tab:novtestathon-ranks}.
As can be seen, TestGen-LLM landed in sixth place in rank order by number of tests generated.
The table also reveals the way in which TestGen-LLM behaves differently to test engineers.
Whereas  test engineers will tend to write a whole class of tests in a single diff, 
TestGen-LLM submits each test as a separate diff.

This is because TestGen-LLM is extending existing test classes, with additional test cases.
It also allows the engineers to more easily accept or reject the recommendations, per test case.
As a result, TestGen-LLM's number of test cases per diff is always one; a fact which  makes it superficially appear to be more productive in terms of diffs.
However, since the ranking was computed in terms of the number of test cases landed, the rank position of TestGen-LLM 
is a fair reflection of its productivity compared to human engineers during the test-a-thon.

Another apparently anomalous result from the table is the number of lines covered by the 17 test cases generated by TestGen-LLM.
At first glance, it might appear that TestGen-LLM is able to cover a great deal more than the human engineers. 
However, this result arose due to a single test case, which achieved 1,326 lines covered. 
This test case managed to `hit the jackpot' in terms of unit test coverage for a single test case.
Because TestGen-LLM is typically adding to the existing coverage, 
and seeking to cover corner cases, the typical expected number of lines of code covered per test case is much lower. 
For example, 6 of the 17 test cases it generated covered only a single extra line. 
However, in all 17  cases, we  manually verified that the generated test did, indeed, cover at least one additional valid  corner case, 
such as an early {\tt return} and/or special processing for special values such as {\tt null} and empty list.
The median number of lines of code added by a TestGen-LLM test in the test-a-thon was 2.5.
This is a more realistic assessment of the expected additional line coverage from a single test generated by TestGen-LLM.


\begin{table}
\begin{tabular}{||r l r r r ||} 
 \hline
rank & test author          &  No, of & lines   &  diffs                            \\
     &                      & tests   & covered &                                   \\
\hline     
1.   & Threads Engineer     &  40                   & 1,047         &  8            \\   
2.   & Home Engineer        &  34                   & 650           &  6            \\   
3.   & Business Engineer    &  34                   & 443           &  3            \\   
4.   & Sharing Engineer     &  33                   & 816           &  8            \\   
5.   & Messaging Engineer   &  18                   & 157           &  2            \\   
6.   & {\bf TestGen-LLM}    & 17                    & 1,460         & 17            \\
7.   & Friends Engineer      & 12                    & 143           & 2             \\   
8.   & Home Engineer        & 10                    & 273           & 2             \\   
9.   & Creators Engineer    & 10                    & 198           & 3             \\   
10.  & Friends Engineer     & 10                    & 196           & 5             \\   
 \hline

\end{tabular}
\caption{Results from the First Instagram Test-a-thon, conducted in November 2023. 
Human test authors are named by the product component on which they worked.
The TestGen-LLM tool landed in sixth place overall, demonstrating its human-competitive added value. }
\label{tab:novtestathon-ranks}
\end{table}

\subsection{Evaluation of LLMs and Prompts}
\label{sec:extend-test}
\label{sec:experiment}
\label{sec:eval}
The outcome of the first November Instagram test-a-thon gave us confidence that we had a usable tool in TestGen-LLM.
In particular, the manual verification that the tests added valid corner cases, the fact that the test style was appreciated by engineers, and the overall performance relative to human effort (see Table~\ref{tab:novtestathon-ranks}), provided the evidence that it was worth developing the tool further, and deploying it more widely.

However, before deploying more widely, we needed to choose suitable defaults for hyper parameters so that engineers could use the tool out-of-the-box without having to consider choices of settings.
To tackle this need we switched TestGen-LLM from deployment mode to evaluation mode.

We undertook experiments to determine the the most favorable parameters among different temperatures, language models available, and prompts.
We conducted experiments on two products for Instagram, Reels and Stories, to determine the differential performance of the two different language models (LLM1 and LLM2), and also to investigate the unique contribution of each of four prompting strategies.
This produced results over 86 Kotlin components with existing human-written test classes (31 for Stories and 55 for Reels) as follows:

\begin{enumerate}
    \item 75\% of test classes had at least one new test case that builds correctly.
    \item 57\% of test classes had at least one test case that builds correctly and passes reliably.
    \item 25\% of test classes had at least one test case that builds correctly, passes and increases line coverage compared to all other test classes that share the same build target.
\end{enumerate}

\begin{figure}[t]
\centerline{\includegraphics[width=0.9\linewidth]{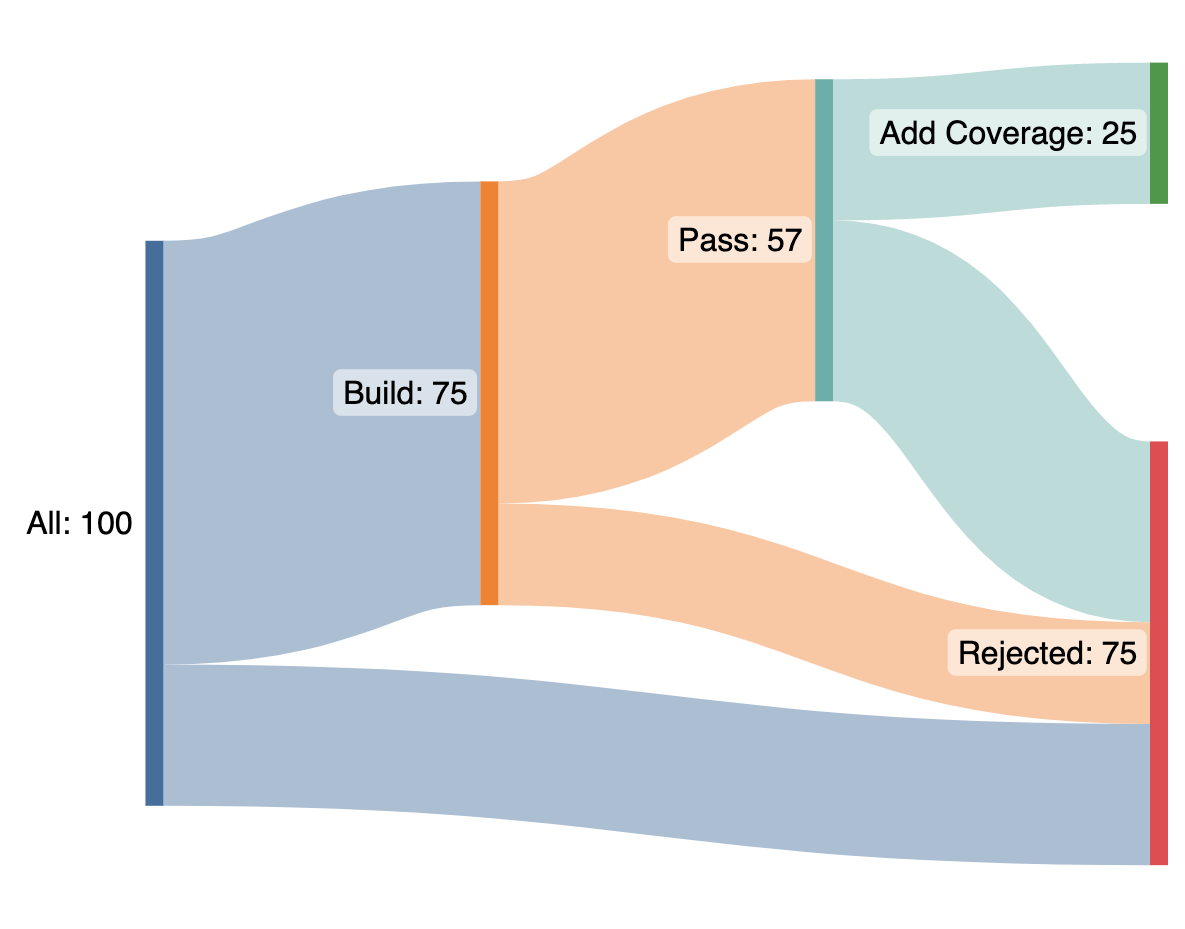}}
\vspace{-3mm}
\caption{Sankey diagram showing the filtration process outcomes (as percentages of all test cases) from the Experimental Study on Instagram  components for Reels and Stories products, 
using the four prompt strategies from Table~\ref{tab:prompts} and the two language models, LLM1 and LLM2. 
\label{fig:sankey}}
\end{figure}

These results are depicted\footnote{This image was created from the data using the freely-available tool SankeyMatic ({\tt https://sankeymatic.com/build/}).} in the Sankey diagram \cite{otto:overview} in Figure~\ref{fig:sankey}.
It was particularly striking that, although 57\% of test classes have a test that builds correctly and passes reliably, only 25\% of classes had a test that builds reliably, passes non-flakily and adds additional line coverage\footnote{Unfortunately, although Jacoco is theoretically capable of collecting branch coverage, this is not available at the scale of testing required by Meta, so we currently rely solely on line coverage.}.

The prompts used are set out in Table~\ref{tab:prompts}.
We wanted to experiment with a variety of different prompting strategies.
The prompt {\tt extend\_coverage}  is the canonical example, which gives maximal information and clear direction to the language model.
{\tt corner\_cases} was included specifically to focus on corner cases, 
while 
{\tt extend\_test} was included to investigate the potential to find solutions when only the test class 
is provided (and not the class under test).
Finally, {\tt statement\_to\_complete} was included to investigate the alternative prompting style of making a statement that should be completed by the language model. 
This is inspired by the fact that  language models are, inherently, predictive models for text completion. 
As such, it seems reasonable that such a prompt ought to have an advantage.

Using these four prompts and the two language models available, we obtained the following results over the 86 test classes:
Using just LLM1, TestGen-LLM was able  to find 
13 tests (15\% of files have at least one new test).
Three of the four prompts added unique value to the overall search:
{\tt extend\_coverage} added 2 unique tests, while 
{\tt corner\_cases} and 
{\tt extend\_test}
each added one unique test.
Although it did succeed in finding test cases,  the prompt 
{\tt statement\_to\_complete} failed to add any {\em unique} test cases, and therefore made no unique individual contribution when using the LLM.

Using just LLM2, TestGen-LLM was able  to find 
16 tests (19\% of files have at least one new test).
All four prompts added unique value to the overall search:
{\tt extend\_coverage} 
added 5 unique tests,
while 
{\tt statement\_to\_complete} 
and 
{\tt extend\_test}  each added 2 unique tests.
{\tt corner\_cases} added 1 unique test.

We also found that generated test cases were most likely to build, pass and extend coverage, with LLM temperature zero. 
However, the effect size was low compared to the nearest next-best temperatures of 0.2 and 0.5; 
all had approximately 4\% success over all test cases suggested\footnote{For a given attempt to extend a test class, there can be many attempts to generate a test case, so the success rate per test case is typically considerably lower than that per test class.}. 
We therefore set the default temperature to zero.

Based on these results, it was  decided to deploy TestGen-LLM for subsequent test-a-thons with default temperature zero, default LLM set to LLM2 and default prompt {\tt extend\_coverage}.
However, users were free  to deploy using other settings when applying the TestGen-LLM tool.
In particular, for subsequent Instagram and Facebook test-a-thons, the tool offered the option to perform a temperature sweep over all temperature settings (from 0.0 to 1.0 in steps of 0.1), 
the option to select a particular prompt and the option to select a particular LLM.
The tool also offered an `LLM ensemble' that combines results from both then-available  LLMs.

\begin{table*}
\begin{tabularx}{\textwidth}{|p{3cm}|X|}
 \hline
Prompt  & Prompt  \\
name    & Template   \\
\hline 
extend\_test            &  Here is a Kotlin unit test class: \{existing\_test\_class\}. 
                           Write an extended version of the test class that includes additional tests to cover some extra corner cases.\\
extend\_coverage        &  Here is a Kotlin unit test class and the class that it tests: 
                           \{existing\_test\_class\} \{class\_under\_test\}. 
                           Write an extended version of the test class that includes additional unit tests that will 
                           increase the test coverage of the class under test.\\
corner\_cases           &  Here is a Kotlin unit test class and the class that it tests: 
                           \{existing\_test\_class\} \{class\_under\_test\}. 
                           Write an extended version of the test class that includes additional unit tests 
                           that will cover corner cases missed by the original and will 
                           increase the test coverage of the class under test.\\
statement\_to\_complete &  Here is a Kotlin class under test \{class\_under\_test\} 
                           This class under test can be tested with this Kotlin unit test class \{existing\_test\_class\}. 
                           Here is an extended version of the unit test class that includes additional unit test 
                           cases that will cover methods, edge cases, corner cases, 
                           and other features of the class under test that were missed by the original unit test class:\\
\hline
\end{tabularx}
\caption{The four primary prompts used in the deployment for the December 2023 Instagram and Facebook app test-a-thons}
\label{tab:prompts}
\end{table*}

\subsection{Deployment in TestGen-LLM-only Instagram Test-a-thon December 2023}
\label{sec:igtestathon2}

During the period 18th to the 20th of December 2023, the (now further improved) TestGen-LLM tool was run on the same directories that were updated in the November test-a-thon. 
We chose these because they have been the target of recent human development so they 
presented a challenge: to improve further on the combination of recent human and previous TestGen-LLM test effort.

Although there was no human intervention in the generation of the diffs by TestGen-LLM, the recommendations were first considered by two of authors before being passed on to engineers.
Engineers were also pre-warned that they may be receiving test cases generated by TestGen-LLM, although there was no other pre-training involved other than providing the context that the diffs they would be receiving were generated by an MVP TestGen-LLM deployment.

This deployment was entirely automated, with TestGen-LLM now running  automatically on these directories without human intervention.
TestGen-LLM automatically generated 42 diffs that were submitted for review. 
Of the 42 diffs:

\begin{itemize}
\item 36 were accepted by the engineer reviewing them.
\item 4 were rejected and/or abandoned.
\item 2 were withdrawn.
\end{itemize}

The 2 withdrawn diffs each added coverage, but we recognized that the files covered were unimportant. 
The reasons for the four rejected diffs were:

\begin{enumerate}
\item Generating tests for trivial methods under test (a getter method).
\item Failing to follow the `single responsibility per test case' principle (2 rejected for this reason).
\item Failing to include an assertion in the test case.
\end{enumerate}

\subsubsection{Deployment at the Facebook App Test-a-thon December 2023}
In this deployment, we had sufficient confidence to automatically submit recommendations from TestGen-LLM to engineers.
There was no engineer pre-training process, no specific test-a-thon expectations, and no additional context provided to the engineers.
This gave us a realistic assessment of the engineers' response to LLM-generated test recommendations provided `out of the box'.
Overall, over 50\% of the diffs submitted were accepted by developers, a figure which rises to almost 70\% of those which received a review by developers.
Specifically, of the 280 diffs generated by TestGen-LLM:

\begin{itemize}
\item 144 were accepted by the engineer reviewing them.
\item 64 were rejected and/or abandoned.
\item 61 did not receive a review.
\item 11 were withdrawn.
\end{itemize}

\section{Quantitative results from deployment}
This section reports the overall results from the fully freestanding deployment at the November and December Instagram and Facebook app test-a-thons.
All deployment data collection took place between 29th Oct 2023 - 29th Dec 2023, during which period TestGen–LLM was deployed (and redeployed) through three different test-a-thons, with  evolution  an improvement in between each iteration.

In total, over the three test-a-thons,
196 test classes were successfully improved, while the  TestGen-LLM tool was applied to a total of 1,979 test classes. 
TestGen-LLM was therefore able to automatically improve approximately 10\% of the test classes to which it was applied. 
Over all, 73\% of TestGen-LLM test improvements were accepted by developers.
This is an encouraging result for an automated code improvement recommender system. 
For example, it compares favorably with previous attempts to deploy automated repair at Meta (where 50\% of recommended fix improvements were accepted and landed into production \cite{ametal:sapfix}).




Table~\ref{tab:platform-ranks} shows the overall success rate, per test case generation trial, over the two platforms: Facebook and Instagram.
In this table, a trial is an attempt to generate a new test case.
A trial is only considered successful if the generated test case builds, passes, and increases coverage over all existing test cases.
As can be seen, the success rate is similar for both platforms, although slightly higher for Facebook than for Instagram.
We believe that this difference may be due to the difference in available training data.
There is approximately one order of magnitude more human-written Kotlin test code for Facebook then there is for Instagram.

\begin{table}
\begin{tabular}{||l | r |  r |  r ||} 
 \hline
Platform      & Successful & Total  & Success \\
              & trials     & trials & rate    \\
\hline              
Facebook      & 490        & 8,996  & 0.05    \\
Instagram     & 831	       & 23,535 & 0.04    \\
 \hline
\end{tabular}
\caption{Results for the two different platforms. The technology was initially developed by the Instagram Product Performance Organisation within Meta, hence its greater number of overall trials. The highest success rate for the Facebook platform may arise from the fact that there are ~10x more examples of human–written test cases for Facebook compared to Instagram. }
\label{tab:platform-ranks}
\end{table}

Table~\ref{tab:temp-ranks} shows the performance over both language models and all four prompts for different temperature settings.
A `successful' trial is one in which a test case is generated that passes all filters 
(successfully builds, passes reliably, and adds additional coverage over all existing tests, including previous LLM-generated tests).
Since the default temperature setting was 0.0, this received by far the greatest number of trials (30,483) during the test-a-thons.
However, as can be seen, other temperature settings were used in the deployment, since engineers applying the tool had the ability to select different temperatures.

It is interesting to note that good results were obtained for a  temperature of 0.4.
It might be tempting to speculate that these results should occasion a change in the default temperature setting.
However, care is required in interpreting such results. 
In particular, the results are based on very different sample sizes.
Furthermore, due to these results being obtained from {\em deployment} mode not {\em experimental} mode, there are unavoidable confounding factors.

The primary comfounding factor is that tests are generated and deployed 
{\em incrementally} in production (as stacks of diffs), 
accumulating coverage as they are deployed.  
Therefore, it becomes incrementally harder to increase coverage over additional production trails.
Experimental mode is a kind of `dry run' in which no tests are actually added and therefore additional coverage is always measured with respect to the same baseline.

Coverage  growth is inherently  logarithmic (over the number of test cases generated).
Therefore,  as more coverage is achieved, it becomes incrementally harder to achieve further improvements from the diminishing number of remaining coverage improvement opportunities.
Therefore, any  configuration setting (such as temperature) that receives a larger share of overall production deployment trials is at a disadvantage compared to those that receive a smaller share.

\begin{table}
\begin{tabular}{||r | r | r | r ||} 
 \hline
Temperature & Successful & Total  & Success \\
            & trials     & trials & rate    \\
\hline 
0.9	        & 50	     & 1,580  & 0.03    \\
0.8	        & 18    	 & 703    &	0.03    \\
0.7	        & 12 	     & 552    & 0.02    \\
0.6	        & 7	         & 536    & 0.01    \\
0.5	        & 4	         & 500    &	0.01    \\
0.4	        & 16	     & 334    &	0.05    \\
0.3	        & 5	         & 324 	  & 0.02    \\
0.2	        & 3	         & 505	  & 0.01    \\
0.1	        & 4	         & 552	  & 0.01    \\
0.0	        & 1,215	     & 30,483 & 0.04    \\	
\hline
\end{tabular}
\caption{Results for different temperature settings. 
After initial experimentation, zero was chosen as the default, hence its far greater number of overall trials. }
\label{tab:temp-ranks}
\end{table}

Finally, we report the results for the two different language models used in the test-a-thons, overall (Table~\ref{tab:llm-ranks}) and per platform (Table~\ref{tab:llm-platform-ranks}).
Since LLM2 was the default model, it received a far greater number of trials and therefore care is required and interpreting the slight differences in performance between the models.
These results do, nevertheless, provide further confirmation of the finding that performance overall is slightly better for the Facebook platform, which enjoys a larger number of existing Kotlin human-written tests, compared to Instagram.

\begin{table}
\begin{tabular}{||l |  r | r | r ||} 
 \hline
LLM           & Successful & Total  & Success \\
              & trials     & trials & rate    \\
\hline              
LLM1         & 163	     & 3,173  &	0.05      \\
LLM2  & 1,157	 & 28,654 & 0.04      \\
 \hline
\end{tabular}
\caption{Results for the two different LLMs. After initial execution LLM2 was chosen as the default, hence its greater number of overall trials. }
\label{tab:llm-ranks}
\end{table}

\begin{table}
\begin{tabular}{||l l | r |r | r ||} 
 \hline
 Platform    & LLM           & Successful & Total  & Success \\
             & used          & trials     & trials & rate    \\
\hline   
Facebook     & LLM1         & 47          &	719	   & 0.07 \\
Instagram    & LLM1         & 116         & 2,454  & 0.05 \\
Facebook     & LLM2         & 443	      &  8,146 & 0.05 \\
Instagram    & LLM2         & 714	      & 20,508 & 0.03 \\
 \hline
\end{tabular}
\caption{Results for the two different platforms and LLMs. }
\label{tab:llm-platform-ranks}
\end{table}


\section{Qualitative observations from deployment}
In this section we present observations and lessons learned from deployment of a more qualitative nature.
These will form the most immediate future work in the technical development of current TestGen-LLM deployment, while
Section~\ref{sec:future} presents higher-level directions for future work and open research problems on which we would be interested in collaborating with and/or learning from the wider research community.

Source code analysis and manipulation have always been, and will likely always remain important in Software  Engineering \cite{mh:scam10-keynote}.
Previous work has successfully used  hybrids of static analysis and language models, in which both applications of static analysis and applications of language models benefited  \cite{mutasim:leveraging,ahmed_improving_2023,li:assisting,jin:inferfix}.
Many of our observations further underscore the potential of static analysis,  and highlight the opportunities for combining static analysis with language model inference.
In particular, we envisage the following four avenues for static analyses to improve LLM inference:

\medskip
\noindent
{\bf
1. LLM `self-plagiarism':}
Since an LLM is, at heart, a probabilistic inference engine, it can be expected that it may produce the same (or similar) responses for the same prompt over multiple samples.  
We observed that both LLM1 and LLM2 often generated almost  identical tests for the same prompt; different in name only. 
This may also be a byproduct of the default temperature setting (zero), which is the most deterministic.

Anecdotally, it seemed that tests generated were either
almost verbatim copies of others previously generated, or very different.
They were never just `somewhat different' on a `sliding scale' of syntactic and semantic difference; the similarity between generated tests was `all or nothing'.

Perhaps the nature of the conditional probability sampling makes this behavior highly likely, and thus expected. 
In later, more mature, versions of TestGen-LLM we doubled down on this observation and included an extra filter to remove previously seen test cases.
The observation that similarity between generated tests was `all or nothing' greatly simplified this filter, because it simply needed to check for syntactic equality of test bodies.
Analyses that detect semantically similar code, such as Type 2 and Type 3 clones \cite{svajlenko:survey} may also be helpful here.

\medskip
\noindent
{\bf
2. Nuanced coverage reporting}: It can happen that TestGen-LLM obtains valuable additional coverage, but not {\em necessarily} solely for the class under test. 
It would be easy to automate the process of identifying this case.
Sometimes this could be very valuable, because it may test parts of the code that are hard to reach in other ways.
Alternatively, where the class under test has little coverage and most of the coverage improvement concerns other units, it may be a sign of inadequate mocking.

TestGen-LLM includes a filter for test flakiness.
This tends to reduce any concerns about inadequate mocking.
Nevertheless, further automated static analysis and post processing can be applied, based on a more nuanced analysis of the coverage achieved.
As the most immediate next step, we plan to flag the issue to code  reviewers, so that they are more  aware of situations in which
generated tests may be playing the role more of integration tests.

\medskip
\noindent
{\bf
3. Highlighting test need:}
Sometimes the generated tests included a "TODO" (e.g., to write the  assertion; see, for example, Section~\ref{sec:cheeky-todo}). 
We did not land these tests into production.  
However, in some cases, they indicated considerable potential coverage wins.
In one case, a series of such tests were generated for the class under test, each of which covered a non-trivial function in the class under test that was not previously covered.

Although these test cases did not include any assertions, they could nevertheless add value simply by covering these functions and using the so-called `implicit oracle' \cite{ebetal:oracle} (that the code should not raise an exception).
Furthermore, such cases might be  useful as hints to human test writers, and may also create the initial template for such a follow-up human-written test.
After all, the task `add a suitable assertion to this generated test' involves considerably less human effort than
the task `write a brand new test case from scratch'.

\medskip
\noindent
{\bf
4. Re-prompting:}
Sometimes, newly-generated tests covered a subset of the lines of the method under test, that had not previously been covered at all. 
This could be a situation where TestGen-LLM should automatically recognize and re-prompt the LLM to try and achieve further coverage for this method. 
If it can: fine. 
If not, TestGen-LLM should flag this situation  to the engineer. 
The human engineer will likely more easily fill in the gaps, now they have  a starting test template to work from.

\section{Related Work}
Software test generation is one of the most widely-studied topics within the more general area of what might be termed `Large Language Model-based Software Engineering' (LLMSE) \cite{mhetal:LLM-survey}.
Wang et al. \cite{wang:llm-test-survey} presented a  literature review  of  102 papers on testing,  debugging and repair,
while Fan et al. \cite{mhetal:LLM-survey} present a general survey, across all software engineering applications, including software testing.

Although both surveys confirm the prevalence of LLM-based test generation in the literature, 
no previous paper  has tackled the problem of extending existing test classes, nor reporting results on the provision of measurable  assurances for both the improvement and the absence of regressions.
The primary technical novelty of the present paper is to introduce this test extension application, as a specific example of Assured LLMSE \cite{mhetal:intense24-keynote}, while the main contribution is the experience report describing its development and deployment at Meta, where it has been applied to Facebook and Instagram.
We believe this is the first instance of industrial deployment of Assured LLMSE.

Results vary quite widely in the literature for coverage achieved by LLM-based test generation.
For example, Siddiq et al.~\cite{siddiq_exploring_2023}
reported that by generating tests from scratch (rather than seeking to improve on existing tests) it
is possible to achieve 80\% coverage on
the small examples in the `HumanEval'  data set using CodeX ~\cite{chen_evaluating_2021-short}.

However, they also report that generation from scratch achieved no more than 2\% coverage on the EvoSuite SF110 data set \cite{fraser:sf110}.
Naturally, one might expect that the coverage achievable would  depend partly on the size of the system, since larger systems have more scope for complex interactions, and deeply nested code that is harder to cover.
Schafer et al.~\cite{schafer_adaptive_2023} also report high statement level coverage (70\%), but also for relatively smaller systems (25 packages, ranging in size from 25 lines of code to 3,100 lines of code).
This indicates the importance of studying and reporting experience from deployment on large complex industrial software systems, 
as a  complement to results on such smaller systems, benchmarks and open source software.

Notwithstanding the high degree of variability for coverage reported in the present literature,
our empirical findings  are broadly consistent  with previous results reported on larger open source systems.
We found that approximately approximately 57\% of Kotlin test cases generated are executable (with 25\%  improving coverage).
This compares relatively favorably with recently reported results.
For example
Nie et al. \cite{nie:learning} report 29\% of tests generated using TeCo are executable,
while
Yuan et al.~\cite{yuan_no_2023}  report approximately one third Chat-GPT-generated  tests are executable with suitable prompt engineering.
However, the language model technologies used, the training set, fine tuning another characteristics play a crucial role \cite{mhetal:LLM-survey}.


TestGen-LLM embeds the language model within a wider software engineering process that filters candidates, in order to provide the assurances that guard against hallucination, and otherwise sub-optimal results, that might accrue from the unfettered application of language models without such filtering.
In this regard, TestGen-LLM is a hybrid approach combining traditional software engineering and software testing with language models as an engine for code generation.
Previous authors have also considered hybrid forms of LLM applications in testing, for example, hybridizing with Search Based Software Testing (SBST) \cite{lemieux_codamosa_nodate}, Mutation Testing \cite{moradi2023effective} and Fuzzing \cite{hu2023augmenting}.

A wide variety of LLMs have been used in previous work on problems relating to software testing, including
BART,
CodeBert,
ChatGPT,
CodeX,
CodeT5,
and
T5 \cite{wang:llm-test-survey,mhetal:LLM-survey}.
Meta released  public versions of LLaMA and CodeLlama, including source code,  in February and August 2023 respectively.
Llama is a general purpose LLM with a variety of model sizes ranging from 7 billion to 65 billion parameters~\cite{touvron:llama}.
CodeLlama is a model more specifically trained on software and has model sizes 7B, 13B, and 34B (and, as of 29th Jan 2024, 70Bn) \cite{codeLlama}.
These two freely available LLMs, LLaMA and CodeLlama, have also been used in previous work on testing topics by other researchers \cite{moon:coffee,xia2023universal}.
Although the results presented here are based on  TestGen-LLM using two internal LLMs built by Meta, the Assured LLMSE design is LLM-agnostic and  allows for an arbitrary number of LLMs to each contribute test cases to the extended test class.

TestGen-LLM's approach to test improvement  draws inspiration from previous research on Genetic Improvement\cite{Petke:gisurvey} and automated repair \cite{legoues:cacm-survey}.
Genetic Improvement  treats existing software code as `genetic material' to be mutated and recombined to improve existing code according to measurable improvement criteria.
Many approaches to automated repair 
adopt a generate-and-test approach, in which multiple candidate solutions are generated using a cheap generation technology, and subsequently filtered and discarded according to evaluation criteria.
Like automated repair, TestGen-LLM uses a generate-and-test approach; filtering out those candidates that do not meet well-defined semantic criteria, such as passing reliably.
Like genetic improvement, TestGen-LLM treats code as genetic material to be mixed and matched. 

However, unlike Genetic Improvement, TestGen-LLM uses an `ensemble' of language models and configurations, rather than genetic programming.
By contrast, in its original formulation \cite{blmh:tec1}, Genetic Improvement envisaged Genetic Programming as the core technology for creating candidate code variations. 
Much of the work on automated program repair has also used similar computational search techniques \cite{legoues:cacm-survey}.
However, the advent of LLMs provides us with an additional route to achieve the same goal.
Essentially, our approach can be thought of as a form of Search Based Software Engineering (SBSE)\cite{mhbj:manifesto,mhamyz:acm-surveys}, in which the search is over a set of candidate test class improvements, which are evaluated using a generate-and-test search process, and for which the core technology for generation is based on language models.

\section{Future Work and Open Problems}
\label{sec:future}
There are many avenues for future work and open problems concerning automated test improvement using Assured LLMSE.
In this section we outline three such open problems.

\medskip
\noindent
{\bf 
1. Assessing improvement}:
The measurement of improvement is clearly a key factor in any Assured LLMSE \cite{mhetal:intense24-keynote}.
We have taken the simple approach of measuring line coverage as a proxy for improvement, but it  is merely  an expedient proxy for `improvement'.

When we use TestGen-LLM in its experimental mode (free from the confounding factors inherent in deployment), 
we found that the success rate per test case was 25\% (See Section~\ref{sec:experiment}).
However, line coverage is a stringent requirement for success.
Were we to relax the requirement to  require only that test cases build and pass, then the success rate rises to 57\%.

Future work will therefore consider other test improvement criteria. 
Mutation coverage \cite{yjmh:analysis} would likely be the best performing criterion.
This is because strong mutation coverage has been empirically demonstrated to outperform other forms of coverage \cite{mike:icse17}.
However, it is challenging to deploy such computationally demanding techniques  at the scale we would require \cite{beller:what}.

\newpage
\noindent
{\bf 
2. Application–aware probability distribution resolution 
}

Language models ultimately produce a conditional probability distribution.
This is typically `resolved' to a single answer using a search algorithm over the distribution, for which the `temperature' parameter is often used to control for variability of outcomes over repeated trials.
Loosely speaking, the temperature determines how `creative' or `exploratory' the overall  process becomes.
Most of the existing work on LLMSE has used default temperature settings  \cite{mhetal:LLM-survey}.
More research is required to define application-specific techniques for transforming the LLM probability distribution into  code \cite{mhetal:intense24-keynote}.

\medskip
\noindent
{\bf 
3. LLMs are dedicated followers of fashion: 
}
LLMs are `fashion followers' that mimic existing test writing styles,  adopting (and often very faithfully replicating) the mode of expression prevalent in the  test class itself and, more generally, in the code-base on which they have been trained. 
This is a natural consequence of the probabilistic nature of the language model.
Often this `fashion following' is a very desirable characteristic.
Feedback from our engineers was very positive regarding the way in which TestGen-LLM tests followed the style of the existing test class.
For example, different components use different assertion styles (standard JUnit style and also bespoke assertion styles, written by engineers as utilities specifically for the component under test).

TestGen-LLM tests follow these styles faithfully. 
In particular, where there was a bespoke style, TestGen-LLM tests use the corresponding utilities.
Furthermore, the generated tests also followed existing naming conventions, commenting styles, and overall test structure.
It would be highly challenging to define algorithms for replicating these bespoke styles using a more rule-based approach, but using an LLM, this useful behaviour simply comes naturally.

Nevertheless, fashion following also means that the language model can pick up deprecated coding habits where these remain prevalent in the code on which the model is trained.
Since training is a periodic and computationally demanding exercise, we cannot simply retrain every time we wish to deprecate a particular style.
In future work we will further augment TestGen-LLM using prompt engineering and additional linter-based filters, and static analysis post processing to address coding style.

\section{Conclusions}
This paper introduced TestGen-LLM which has been used to land test cases in production at Meta.
The paper described the evolution of TestGen-LLM from proof of concept, through minimal viable product, to deployed test support tool.
The primary TestGen-LLM characteristic of interest is the way in which TestGen-LLM guards against LLM hallucination: it  submits, for human review, only test cases that it can guarantee improve on the existing code base.
We believe this is the first report of Assured Large Language Model Software Engineering deployed at scale in industry.

\bigskip

\noindent {\bf Acknowledgements:}
\small
We wish to thank the many AI and developer infrastructure teams at Meta for their work on the LLMs, build, test and continuous integration systems, without which TestGen-LLM would not be possible.
We also want to thank the Instagram and Facebook platform organizations, and leadership for their support, and the many Meta engineers who reviewed the code produced by TestGen-LLM, providing valuable feedback and insights on its development, deployment and evolution. 
\normalsize

\newpage
\balance

%
\bibliographystyle{ACM-Reference-Format}
\bibliography{main}

\end{document}